\documentclass{ws-mpla}

\begin{document}

\newcommand{\nnprime}{n,n$^\prime \gamma$}
\newcommand{\ntwon}{n,2n$\gamma$}
\newcommand{\nthreen}{n,3n$\gamma$}
\newcommand{\nxn}{n,xn$\gamma$}
\newcommand{\nx}{n,x$\gamma$}

\newcommand{\natpb}{$^{\textrm{nat}}$Pb}
\newcommand{\natge}{$^{\textrm{nat}}$Ge}
\newcommand{\enrge}{$^{\textrm{enr}}$Ge}
\newcommand{\eigpb}{$^{208}$Pb}
\newcommand{\sevpb}{$^{207}$Pb}
\newcommand{\sixpb}{$^{206}$Pb}
\newcommand{\fivpb}{$^{205}$Pb}
\newcommand{\foupb}{$^{204}$Pb}
\newcommand{\nonubb}  {$0 \nu \beta \beta$}
\newcommand{\twonubb} {$2 \nu \beta \beta$}
\newcommand{\gam}{$\gamma$}
\def\nuc#1#2{${}^{#1}$#2}
\def\mee{$\langle m_{\beta\beta} \rangle$}
\def\mnu{$\langle m_{\nu} \rangle$}
\def\ml{$m_{lightest}$}
\def\gnu{$\langle g_{\nu,\chi}\rangle$}
\def\mmod{$\| \langle m_{\beta\beta} \rangle \|$}
\def\mb{$\langle m_{\beta} \rangle$}
\def\BBz{$0 \nu \beta \beta$}
\def\BBm{$\beta\beta(0\nu,\chi)$}
\def\BBt{$2 \nu \beta \beta$}
\def\nonubb{$0 \nu \beta \beta$}
\def\twonubb{$2 \nu \beta \beta$}
\def\BB{$\beta\beta$}
\def\Gz{$G^{0\nu}$}
\def\Mz{$M_{0\nu}$}
\def\Mt{$M_{2\nu}$}
\def\MzG{$M^{GT}_{0\nu}$}           
\def\MzF{$M^{F}_{0\nu}$}                
\def\MtG{$M^{GT}_{2\nu}$}           
\def\MtF{$M^{F}_{2\nu}$}                
\def\Tz{$T^{0\nu}_{1/2}$}
\def\Tt{$T^{2\nu}_{1/2}$}
\def\Tc{$T^{0\nu\,\chi}_{1/2}$}
\def\Rz{$\Gamma_{0\nu}$}            
\def\Rt{$\Gamma_{2\nu}$}            
\def\ms{$\delta m_{\rm sol}^{2}$}
\def\ma{$\delta m_{\rm atm}^{2}$}
\def\ts{$\theta_{\rm sol}$}
\def\ta{$\theta_{\rm atm}$}
\def\tot{$\theta_{13}$}
\def\gpp{$g_{pp}$}                  
\def\qval{$Q_{\beta\beta}$}                 
\def\MJ{{\sc Majorana}}             
\def\DEM{{\sc Demonstrator}}             
\def\be{\begin{equation}}
\def\ee{\end{equation}}
\def\cpRty{counts/ROI/t-y}
\def\onecpRty{1~count/ROI/t-y}
\def\fourcpRty{4~counts/ROI/t-y}
\def\ppc{P-PC}                          
\def\nsc{N-SC}

\markboth{Steven R. Elliott}
{Double Beta Decay}

\catchline{}{}{}{}{}

\title{Recent Progress in Double Beta Decay}

\author{\footnotesize Steven R. Elliott}

\address{Physics Division, Los Alamos National Laboratory, Los Alamos, New Mexico,  87545 USA.\\
elliotts@lanl.gov}

\maketitle

\pub{Received (Day Month Year)}{Revised (Day Month Year)}

\begin{abstract}
At least one neutrino has a mass of about 50 meV or larger. However, the absolute mass scale for the neutrino remains unknown. Studies of double beta decay offer hope for determining the absolute mass scale. Furthermore, the critical question: Is the neutrino its own antiparticle? is unanswered. In particular, zero-neutrino double beta decay (\BBz) can address the issues of lepton number conservation, the particle-antiparticle nature of the neutrino, and its mass. A summary of the recent progress in \BBz, and the related technologies will be discussed in the context of the future \BBz\ program.

\keywords{neutrinoless decay, massive Majorana neutrinos, matrix elements, experimental search for \BBz\ decay}
\end{abstract}

\ccode{PACS Nos.: 11.30.Fs, 14.60.Pq, 23.40.-s}

\section{Introduction}
Two-neutrino double beta decay (\BBt) can be described as 2 neutrons simultaneously beta decaying within a nucleus emitting 2 $\beta$ particles and 2 $\overline{\nu}$s. If the neutrino has certain characteristics, the alternative process of zero-neutrino double beta decay (\BBz) may occur where the neutrino is exchanged as a virtual particle between two neutrons and only $\beta$ particles are emitted in the final state. Understanding neutrino properties motivates the study of this process.

At least one neutrino has a mass of about 50 meV or larger. However, the absolute mass scale for the neutrino remains unknown. Furthermore, the critical question ``Is the neutrino its own antiparticle?" is unanswered. Studies of \BBz\ can address the issues of lepton number conservation, the particle-antiparticle nature of the neutrino, and the neutrino mass scale. Recent experimental results have demonstrated the increasing reach of the technologies used to search for \BBz. In addition, theoretical progress in understanding the nuclear physics involved has also been impressive. All indications are that upcoming generations of \BBz\ experiments will be sensitive to neutrino masses in the exciting range below 50 meV. 

The half-life of \BBz\ is directly related to the neutrino mass. But the half life is very long; at least greater than 10$^{25}$ years. Hence any search for the rare peak in a spectrum resulting from \BBz\ must minimize the background of other processes that may take place in a detector. In addition, deducing a neutrino mass value from a half-life measurement or limit requires an understanding of the transition matrix element, which is technically difficult to calculate. 

In this article, a summary of the the related nuclear physics of \BB\ will be discussed in the context of the future \BBz\ program. Numerous reviews have been written on this topic and provide great detail on this exciting science program. (See for example Refs.~\refcite{ell02,ell04,Avi08,Rode11,Bar11,GomezCadenas11}.)

\section{\BBz\ and Neutrino Mass}
The decay rate for \BBz\ can be written:

\begin{equation}
\label{eq:rate}
[T^{0\nu}_{1/2}]^{-1} = G_{0\nu} |M_{0\nu}|^2 \langle m_{\beta\beta} \rangle^2
\end{equation}

\noindent where $T^{0\nu}_{1/2}$ is the half-life of the decay, \Gz\ is the kinematic phase space factor, \Mz\ is the matrix element corresponding to the \BBz\ transition, and \mee\ is the effective Majorana neutrino mass. \Gz\ contains the kinematic information about the final state particles, and is  calculable to the precision of the input parameters (though use of different nuclear radius values in the scaling factors of \Gz\ and \Mz\ have previously introduced some confusion\cite{cow06}). 

Cosmology is sensitive to the sum of the neutrino mass eigenstates ($\Sigma$) and $\beta$ decay endpoint measurements determine a different combination ($\langle m_{\beta} \rangle$) of mass eigenvalues and neutrino mixing parameters than \BBz. The three techniques, therefore, provide complementary information on neutrino physics. The three equations are given by:

\begin{eqnarray}
\label{eq:mass}
\mbox{\mee} &=& \left| \sum_j m_j U_{ej}^2 \right| = \left| m_1 |U_{e1}|^2 + m_2 |U_{e2}|^2 e^{i\phi_1} + m_3 |U_{e3}|^2 e^{i\phi_2} \right|
\\
\langle m_{\beta} \rangle &=& \sqrt{\sum_j m_j^2 U_{ej}^2}  \\
\Sigma &=& \sum_j m_j	
\end{eqnarray}

\noindent where $m_j$ are the neutrino mass eigenstates and $U_{ej}$ are the neutrino mixing matrix elements. Equation 2 is written for 3 light-mass neutrinos. The CP violating phases denoted by $\phi_1$ and $\phi_2$ are functions of the mixing matrix phase and the Majorana phases. (See Ref.~\refcite{ell04} for example.) Present cosmological constraints on $\Sigma$ are in the range 440-760 meV\cite{Hannestad2010,Gonzalez2010} and upcoming measurements hope to reach a sensitivity below 100 meV\cite{Hannestad2006,Lesgourgues2006}. The KATRIN tritium $\beta$-decay experiment hopes to reach a sensitivity for \mb\ of 200 meV\cite{Robertson2008}. $\beta$-decay neutrino-mass results would have the least model dependency of the 3 techniques.  \BBz\ results will be the most sensitive laboratory measurements of neutrino mass if the neutrino is Majorana. The interpretation of cosmology results would be greatly enhanced by a laboratory neutrino mass result with which to constrain models.

An open question in neutrino physics is whether or not the lightest neutrino mass eigenstate is the dominant component of the electron neutrino. If so, we refer to the neutrino mass spectrum as being {\em normal}. If not, we refer to it as {\em inverted}. Figure~\ref{fig:BBmass} shows the effective Majorana neutrino mass as a function of the lightest neutrino mass for these 2 possibilities using the neutrino oscillation parameters from Ref.~\refcite{Fog11}. 

\section{Cancellation Effects of CP Phases}

\begin{figure}[ht]
\includegraphics[width=14pc]{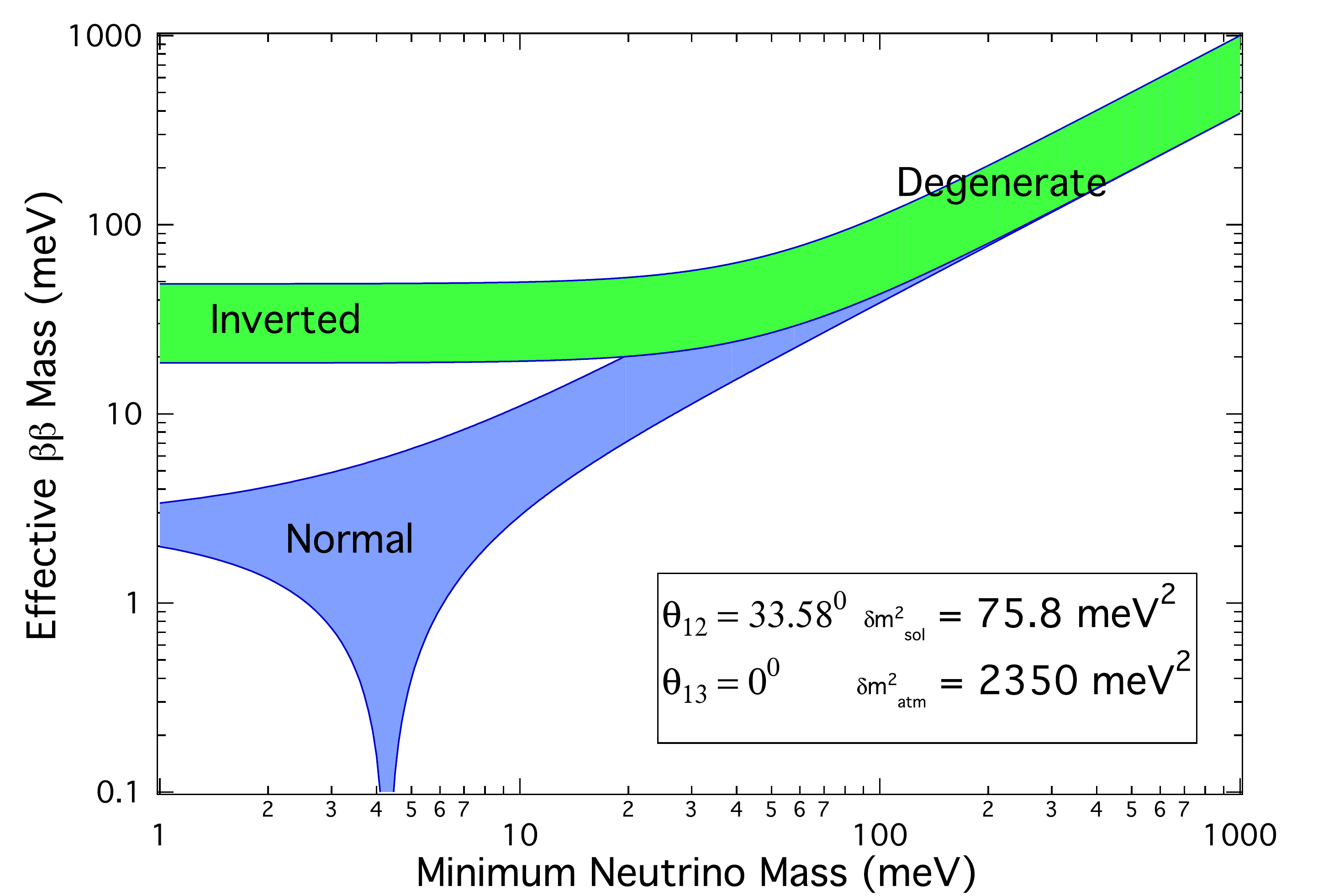}\hspace{2pc}%
\includegraphics[width=14pc]{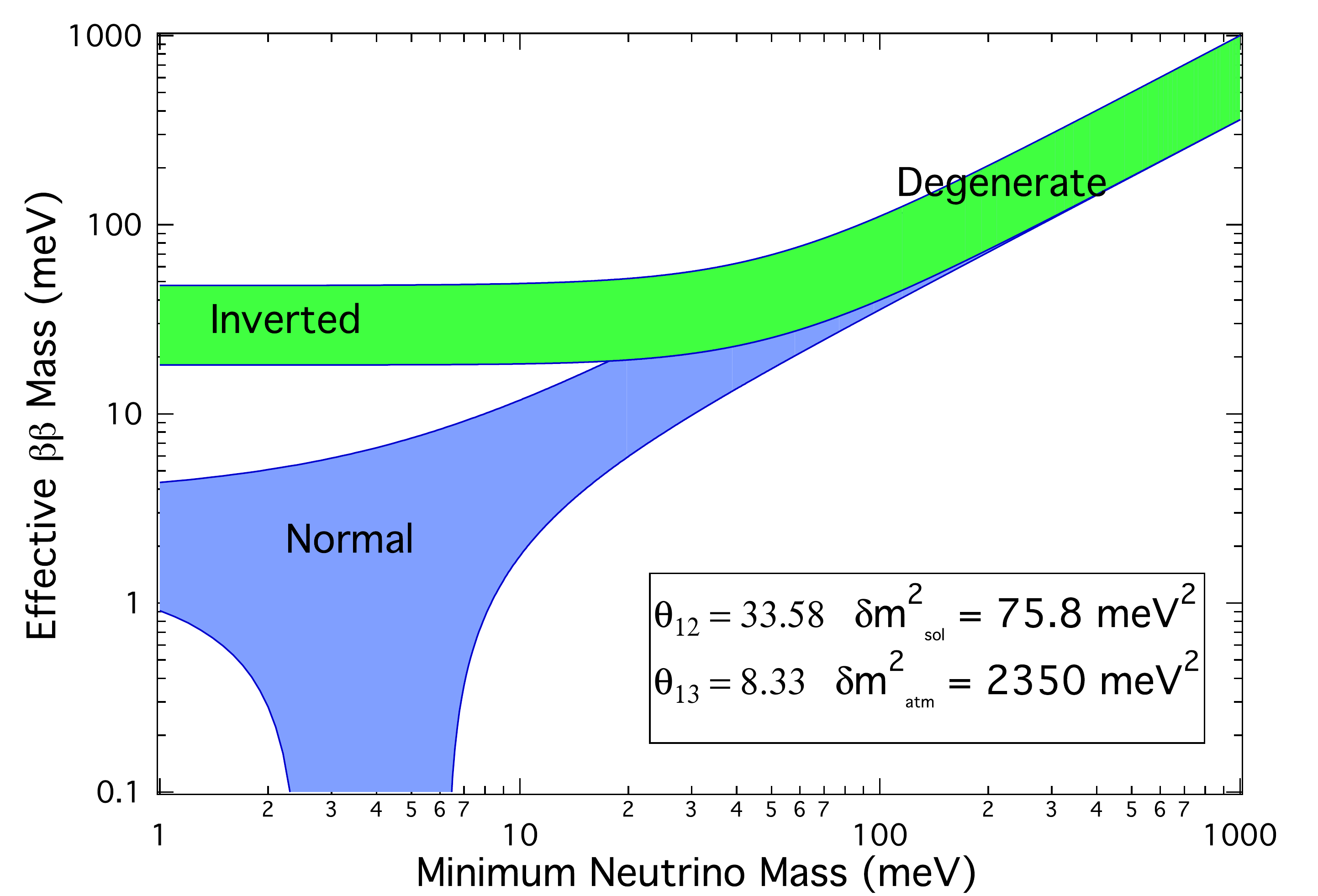}\hspace{2pc}%

\caption{The effective Majorana neutrino mass as a function of the lightest neutrino mass. In the left panel, $\theta_{13}$ was taken to be zero, whereas in the right panel it is set to the best fit value in recent global fits.\protect\label{fig:BBmass}}
\end{figure}

Figure~\ref{fig:BBmass} seems to indicate that a large fraction of the potential parameter space within the normal hierarchy can result in a negligible \mee\ even if neutrinos are Majorana particles. This is a bit misleading because for the expression in Eqn.~\ref{eq:mass} to result in a small \mee, specific values of the mixing elements, mass eigenstates and phases must conspire to cancel. Barring some symmetry that requires such a cancellation, this would be a unnatural coincidence. In fact the fraction of the parameter space that would result in a cancellation is rather small if the parameter values are random. In Fig.~\ref{fig:Contour} one sees that for a given value of $m_1$ in the region of parameter space that can potentially suffer such cancelations, about 5\% of the $\phi_1$-$\phi_2$ space results in \mee\ less than 1 meV.

\begin{figure}[ht]
\includegraphics[width=14pc]{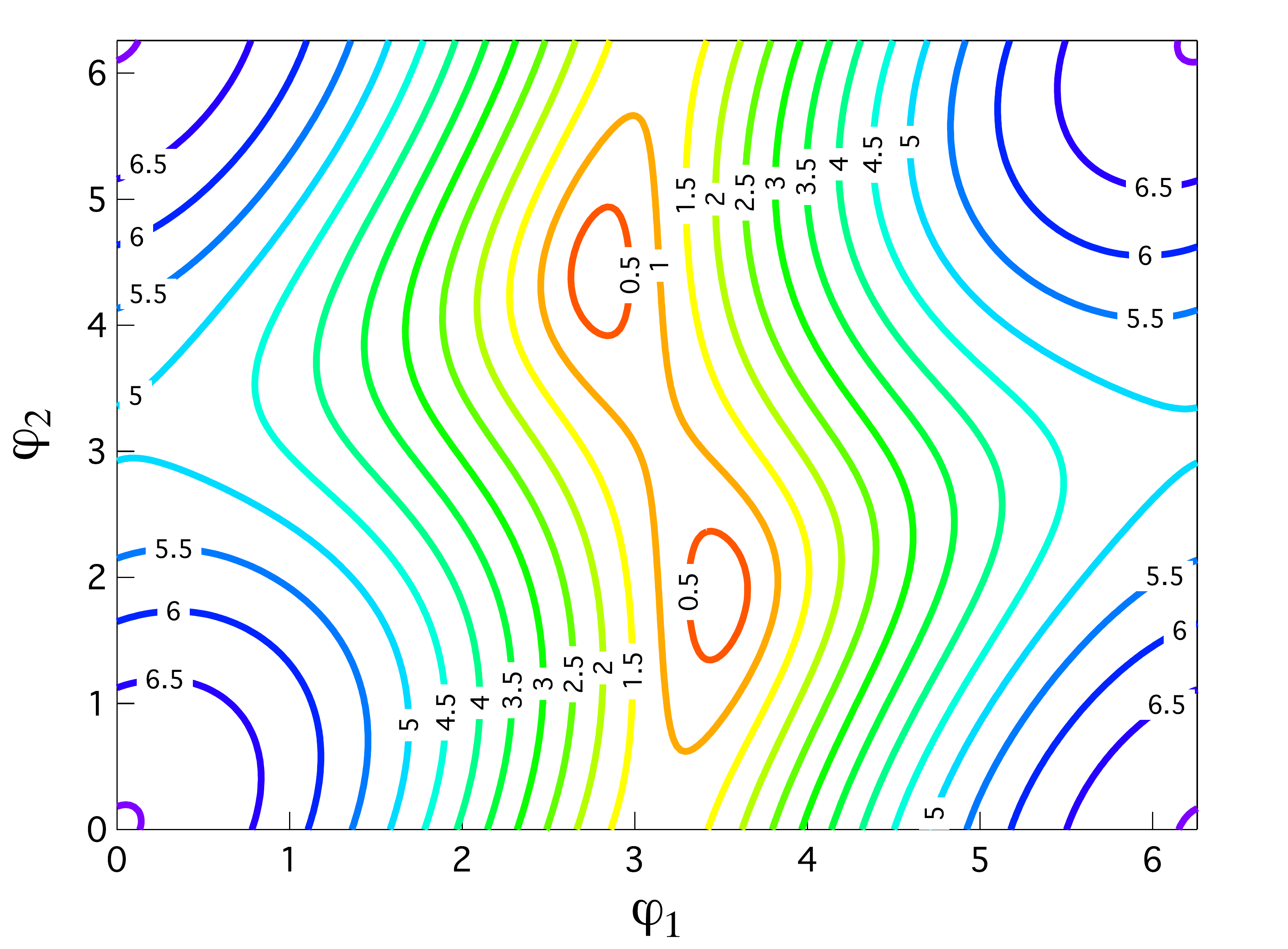}\hspace{2pc}%
\includegraphics[width=14pc]{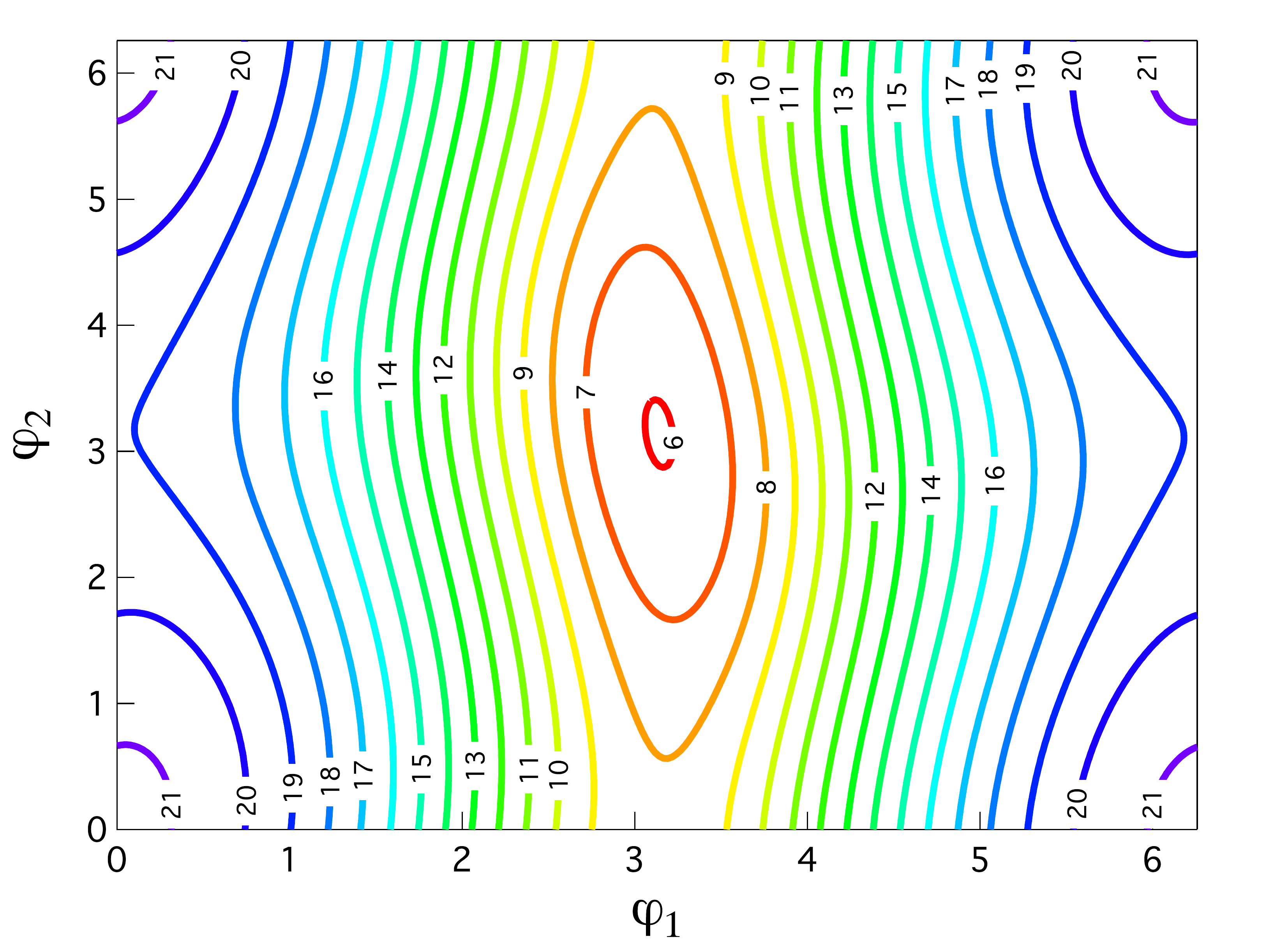}\hspace{2pc}%
\protect\label{fig:Contour}
\caption{Contour plots of \mee\ as a function of the CP violating phases. The mixing angle values were chosen as in recent global analyses. In the left panel $m_1$ was chosen to be 4.5 meV as that resides near the center of the normal hierarchy branch of the \mee\ region that can be subject to significant cancellations for specific parameter values. In the right panel, $m_1$ was chosen to be 20 meV, where no cancelation is anticipated.\protect\label{fig:Contour}}
\end{figure}

\begin{table}[h]
\tbl{A list of recent  \BBz\ experiments and their 90\% confidence level (except as noted) limits on \Tz. The \mee\ limits are those quoted by the authors using the \Mz\ of their choice.}
{\begin{tabular}{@{}ccccc@{}} 
\toprule
Isotope                &  Technique                                             & \Tz\         & \mee\ (eV)       & Reference  \\
\colrule
\nuc{48}{Ca}     & CaF$_2$ scint. crystals             &$>1.4 \times 10^{22}$ y               &   $<$7.2-44.7   & \refcite{oga04}\\
  \nuc{76}{Ge}   & \nuc{enr}{Ge} det.                       &    $>1.9 \times 10^{25}$ y          & $<$ 0.35           & \refcite{kla01a} \\
  \nuc{76}{Ge}   & \nuc{enr}{Ge} det.                       &    $(1.19^{+2.99}_{-0.50}) \times 10^{25}$ y (3$\sigma$)   & 0.24-0.58 & \refcite{kla04a}\\
  \nuc{76}{Ge}   & \nuc{enr}{Ge} det.                       &    $>1.57 \times 10^{25}$ y            & $<$(0.33-1.35) & \refcite{Aal02a} \\
  \nuc{82}{Se}    &Thin metal foils and tracking               & $>3.6 \times 10^{23}$ y              & $<$(0.89-2.54)        &  \refcite{Barabash2011}\\
  \nuc{96}{Zr}	&Thin metal foils and tracking               & $>9.2 \times 10^{21}$ y				&$<$(7.2-19.5)	 & \refcite{Arg10} \\
\nuc{100}{Mo}   &Thin metal foils and tracking               & $>1.1 \times 10^{24}$ y             & $< $(0.45-0.93)         &  \refcite{Barabash2011}\\
\nuc{116}{Cd}   &\nuc{116}{Cd}WO$_4$ scint. crystals  & $>1.7 \times 10^{23}$ y              & $<$1.7       & \refcite{dane03}\\
\nuc{128}{Te}    & geochemical                                         & $>7.7 \times 10^{24}$ y               & $<$(1.1-1.5)         & \refcite{ber93} \\
\nuc{130}{Te}    & TeO$_2$ bolometers                     & $>2.8 \times 10^{24}$ y              & $<$(0.3-0.7)        &  \refcite{arn08}  \\
\nuc{136}{Xe}   &  Xe disolved in liq. scint.  & $>5.7 \times 10^{24}$ y &  $<$(0.3-0.6) & \refcite{Gando2012}    \\
\nuc{150}{Ne}   &   Thin metal foil within TPC    &$>1.8 \times 10^{22}$ y               & N.A.                   & \refcite{Arg09} \\
\botrule
\end{tabular} }
\label{tab:PastExperiments}
\end{table}%

\section{Nuclear Physics and \BB}
The observation of \BBz\ would have profound qualitative physics conclusions.  However, the interpretation of those results quantitatively requires knowledge of \Mz.  Furthermore, an accurate knowledge of \Mz\ has benefits for experimental design. Most nuclear matrix element calculations involve either the quasiparticle random phase approximate (QRPA) technique or the nuclear shell model (NSM).  Although the two methods have a similar starting point (a Slater determinant of independent particles), they are complementary in their treatment of correlations. QRPA uses a large number of ÒactiveÓ nucleons in a large space but with a specific type of correlation suited for collective motion.  NSM uses a small number of nucleons within a limited space but with arbitrary correlations. Three additional techniques have recently been used to estimate \Mz. These include the interacting boson model (IBM-2)\cite{Bare09}, the projected Hartree-Fock-Bogolyubov (PHFB)\cite{Chan09}, and the Energy Density Functional (EDF)\cite{Rod10} techniques.

Recent publications\cite{rod03,Rod06,suh05,Kor07,Fae08,Sim09a,sim09} have elucidated the causes of the historical disparity of results from QRPA calculations of \Mz.  As a result, the technique now provides a uniform result and Ref.~\refcite{sim09} summarizes the values for several isotopes.  The NSM has also seen a resurgence of activity with studies of input micro-physics and its influence on \Mz \cite{Cau08,Cau08a} with recent values given in Ref.~\refcite{Men09}.
This  progress indicates that the agreement between the various calculations is improving, but further work is needed to reduce the uncertainty to levels required for comparisons between \BBz\ results from different isotopes\cite{Geh07,Dep07}.   

A previous workshop\cite{zub05} was dedicated to nuclear physics studies that would support the understanding of \BB. Much work has been done since that workshop, not only on ideas presented there but others that have arisen since. Below we focus on the impact of some of these related nuclear physics measurements.

\subsection{Atomic Masses}
One key aspect of \BBz\ searches is that the energy of the transition is known because the Q-value is deduced from the nuclear masses. The Q-values for many \BB\ transitions, however, were previously not well known. In fact in a number of cases, experimental resolution was better than the uncertainty in the endpoint energy. Recent Penning-trap measurements\cite{Rahaman2008,red07} have been providing a wealth of data to eliminate this uncertainty.  In the case of \nuc{130}{Te}, a new Q-value measurement\cite{red09} resulted in an approximate 5\% \Tz\ limit change due to a shift in the endpoint energy at which the experiment searched\cite{arn08,And11} indicating the importance of the Q-value. For experiments trying to observe resonant radiative zero-neutrino double electron capture\cite{suj04,wyc04} ($0\nu 2EC$), knowledge of the Q-value is critical in order to choose isotopes for which the resonance is possible. A number of these isotope masses have been measured recently also, but no strong resonance has yet been identified. Reference \refcite{Eliseev2011} provides a summary of mass measurements applicable to $0\nu 2EC$ and lists a number of remaining possible candidate isotopes for which a resonance may exist.

\subsection{Neutron Reactions}
Neutron reactions can result in background for \BB\ but the cross sections are often unavailable and require measurement in order to predict the impact on an experiment. For example, the \nuc{207}{Pb}(n,n'$\gamma$) reaction can produce a 3062-keV $\gamma$ ray\cite{mei08}. This is a dangerous background for \BBz\ using \nuc{76}{Ge} because the double escape peak energy coincides with the Q-Value. Many of the cross sections for such reactions, including this example\cite{gui09} were not well known and measurements are required. Reactions such as \nuc{76}{Ge}(n,9n)\nuc{68}{Ge} are also dangerous\cite{Ell10} as they result from cosmic-ray induced neutrons while materials reside on the Earth's surface. The cross sections for such large $\Delta A$ transitions are small however, and many are also frequently not yet measured.

\subsection{Transfer Reactions and Pair Correlation Studies}
Proton- and neutron-removing transfer reactions have played a critical role in providing tests of the nuclear structure used for the calculation of \Mz. The difference in the nucleon configuration of the initial and final nuclei is an important input to \Mz. Reactions such as (d,p), (p,d), ($\alpha$,\nuc{3}{He}), and (\nuc{3}{He},$\alpha$) were used to study the occupation of valence neutron orbits in \nuc{76}{Ge} and \nuc{76}{Se}\cite{Schif08}. For protons, (d,\nuc{3}{He}) reactions were studied on these isotopes\cite{Kay09}. The correlation of pairs of neutrons was measured in this system using the (p,t) reaction\cite{Fre07}. These data were then used to constrain calculations of \Mz\ by both the NSM\cite{Men09a} and QRPA techniques\cite{sim09,suh08}. As a result, the matrix element calculations of both techniques shifted such that the difference between them was reduced. Studies on other \BB\ isotopes are underway.

\subsection{Precise \BBt\ Data}
Accurate half-lives for \BBt\cite{rod03} along with $\beta^{\pm}$ and electron capture decay data of the intermediate nuclei\cite{suh05} can help determine the \gpp\ parameter used in QRPA calculations. Being able to calculate \BBt\ rates can be also considered a necessary, if not sufficient condition for calculating \BBz\ rates. Reference~\refcite{Bar10a} summarizes \BBt\ measurements and compares to the \Mt\ deduced. In addition the recent \BBt\ measurements of \nuc{136}{Xe}\cite{Ack11,Gando2012} will impact future analyses of \BBz.

\subsection{Other Nuclear Physics Measurements for the Study of \Mz}
A variety of other proposals for studying the physics related to \Mz\ have been made and we summarize them here.

{\bf Charge exchange reactions on parent \& daughter:} Charge exchange reactions, such as (p,n), (n,p), ($^3$He,t), etc., can provide data on the Gammov-Teller transition strengths of interest for \BB\cite{eji00a}. The Fermi part of the total \BB-decay nuclear matrix element might be studied in charge-exchange reactions\cite{Rod09}.

{\bf Muon capture:} The \BB\ virtual transition proceeds through levels of  the intermediate nucleus. For \BBt\ intermediate 1$^+$ states are involved, whereas for \BBz\ all J$^+$ states participate. Theoretical calculation of the relative strengths of these virtual states is very difficult. Muon capture on the final nucleus\cite{kor02} also excites all multipoles and therefore provides  additional experimental data with which to compare calculation techniques.

{\bf Neutrino Cross Sections:} Matrix element studies could be done with neutrino beams\cite{vol05}. By varying the average neutrino beam energy, specific multipoles of the intermediate nucleus can be excited. The strengths of both legs of the \BB\ transition could be studied by using both $\bar{\nu_e}$ and $\nu_e$.

{\bf Electromagnetic Transitions to Isobaric Analogue States:} The matrix element for the $\beta$ decay of an intermediate nucleus can be measured by observing the ($\gamma$,p) through isobaric-analog-state excitation on the daughter nucleus\cite{eji06a}.

\section{Upcoming Experiments}
\label{sec:expt}
Within the next 3-5 years, a number of new experiments will begin to provide data on \BBz. All of these programs have a chance to test a recent claim of the observation of \BBz\cite{kla06}. The validity of this claim has been debated\cite{aal02b,fer02,har02,zde02a,Kir10} and therefore confirmation is required. Table~\ref{tab:FutureExperiments} summarizes these upcoming experiments along with various longer term efforts. These projects are complementary in that some emphasize isotope mass (EXO-200, KamLAND-Zen, SNO+), others energy resolution (CUORE, GERDA, \MJ), and still others include precision tracking (NEXT, SuperNEMO). All can potentially lead to technologies with enough isotope mass to reach sensitivity to the inverted hierarchy neutrino mass scale. It is anticipated that the sensitivity of these projects will result in mass limits of 100 meV or less by about mid decade. 

Here we summarize a number of the experiments, however, there are a great number of other R\&D projects that are too numerous to discuss in detail here, but the key points of each are included in Table~\ref{tab:FutureExperiments}.

\subsection{CUORE} 
 CUORE  (Cryogenic Underground Observatory for Rare Events)\cite{Aless11}, and its planned prototype CUORE-0\cite{Aless11}, use TeO$_2$ crystals operated as bolometers. This program builds on the previously successful CUORICINO experiment\cite{And11}. The project exploits the high natural abundance (33.8\%) of \nuc{130}{Te} as the \BB\ isotope. The planned use of 988 crystals provides a total of 203 kg of \nuc{130}{Te}. These bolometers operate at low temperature ($\approx$8-10 mK) where the heat capacity is very small. Therefore an energy deposit due to \BB\ can be measured as a temperature rise in the detector. Energy resolution of 5 keV at the 2528-keV endpoint has been achieved. CUORE-0 and CUORE are being assembled at Gran Sasso and CUORE-0 expects to start data taking early 2012 with the first tower of crystals (10 kg \nuc{130}{Te}).

\subsection{EXO-200} 
 EXO-200 (Enriched Xenon Observatory) is operating at the Waste Isolation Pilot Plant (WIPP). The experiment consists of 200 kg of Xe enriched to 80.6\% in \nuc{136}{Xe} housed in a liquid time projection chamber. This detector measures energy through both ionization and scintillation and is capable of effectively rejecting $\gamma$ rays through topological cuts. Encouragingly, EXO-200 has recently claimed the first observation of \BBt\ in \nuc{136}{Xe}\cite{Ack11}. Initial results on \BBz\ should be forthcoming. The project is also a prototype for a planned 1-10 tonne sized experiment\cite{dan00a} that may include the ability to identify the \BB\ daughter \nuc{136}{Ba} in real time, effectively eliminating all classes of background except that due to \BBt.

\subsection{GERDA} 
GERDA (GERmanium Detector Array) started physics data taking November 2011 at Gran Sasso\cite{sch05}. This project uses high purity Ge detectors, which provide the best energy resolution of any double beta decay experiment. In their Phase-I configuration, they immersed the enriched detectors used by the IGEX\cite{Aal02a} and Heidelberg-Moscow\cite{bau99} collaborations bare into a large volume of liquid argon. The liquid argon vessel is itself enclosed within a large water tank. The Ge detectors total 17.66 kg of Ge metal which is enriched to 86\% in \nuc{76}{Ge}. Initial results should be available soon. For their Phase-II configuration, they have obtained additional enriched material and plan to add an additional $\approx$20 kg of p-type detectors with small read out electrode\cite{dusa09,Barnabe2010,Agostini2011} and begin their commissioning in late 2012 or early 2013.  

\subsection{KamLAND-Zen} 
KamLAND-Zen\cite{Gando2012} (Kamioka Liquid Anti-Neutrino Detector, ZEro Neutrino double beta decay) is an extension of the KamLAND\cite{Abe08} experiment. KamLAND is a 6.5-m radius balloon filled with 1000 t of liquid scintillator, submerged inside a 9-m radius stainless-steel sphere filled with 3000 t of mineral oil with PMTs mounted on the wall. The cavity outside this sphere is filled with water also instrumented with PMTs. KamLAND was built to search for reactor anti-neutrinos and the extension is intended as a search for \BBz. The collaboration added an additional low-background miniballoon into the inner sphere that contains 13 t of liquid scintillator loaded with 330 kg of dissolved Xe gas enriched to 91\% in \nuc{136}{Xe}. This detector at the Kamioka mine in Japan began \BB\ operation in September 2011 and initial results include an improved limit on \Tz\ for \nuc{136}{Xe} and a measurement of \Tt\ that agrees with the recent EXO-200 result.

\subsection{M{\small{AJORANA}} }
The \MJ\ \DEM\cite{Aguayo2011,Phillips2011,Schubert2011} is being constructed as a prototype to show that building a large mass Ge detector with low-enough background to study the inverted hierarchy neutrino-mass scale is feasible. The project consists of 40 kg of Ge p-type point-contact detectors\cite{barb07,Aguayo2011} enclosed within 2 separate electroformed copper cryostats. Up to 30 kg of these detectors will be fabricated from Ge enriched to 86\% in \nuc{76}{Ge}. The cryostats will be contained within a shield that consists of electroformed Cu, OFHC Cu, lead, cosmic ray veto system, and plastic. Electroformed Cu is used due to its extreme purity\cite{Phillips2011,Hoppe2008}. The experiment is being constructed at the Sanford Underground Research Facility (SURF) at the old Homestake gold mine in Lead, SD. The first cryostat is planned for commissioning in 2013. 

The \MJ\ and GERDA collaborations are cooperating in efforts to design a large-mass Ge detector. The configuration of such an experiment will be optimized based on the outcomes of the \DEM\ and GERDA Phase-II.

\subsection{NEXT} 
NEXT\cite{Gom11,Yah10} (Neutrino Experiment with Xenon TPC) intends to use $>$100 kg of Xe enriched to $\sim$90\% in \nuc{136}{Xe}. The detector will be a moderate-density gas TPC $\sim$0.08 g/cm$^3$ that will also detect scintillation light. By operating at low pressures ($\approx$15 bar), the design should not only provide good energy resolution, but also permit tracking that allows fairly detailed track reconstruction to confirm that candidate events involve two electrons moving in opposite directions. The collaboration has recently demonstrated impressive 1\% FWHM resolution in a limited fiducial volume device. Construction is scheduled to start in 2012 with commissioning to start in 2014. It will operate at the Laboratorio Subterr\'{a}neo de Canfranc in Spain (LSC).

\subsection{SNO+} 
SNO+\cite{che05,Chen2008} is an outgrowth of the SNO (Sudbury Neutrino Observatory) experiment that observed solar neutrinos\cite{ahm01}. For SNO+, the heavy water within the large acrylic vessel is replaced by scintillator with dissolved Nd. For a fractional loading of 0.1\% by weight, they will use 43.6 kg of \nuc{150}{Nd}, although higher loadings are being studied. In addition to \BB\ the experiment can also study pep and CNO solar neutrinos along with reactor and geo anti-neutrinos. The experiment is being constructed at SNOLAB in Canada and is expected to begin data taking in mid 2013 with scintillator and with added Nd soon thereafter.

\subsection{SuperNEMO} 
The SuperNEMO\cite{Arn10} proposal builds on the great success of the NEMO-3 (Neutrino Ettore Majorana Observatory) experiment, which measured \BBt\ rates in 7 isotopes\cite{Bon11}. NEMO-3 has provided the best \BBt\ data to date including information on single-electron energy distributions and opening angles. The design uses calorimetry to measure energies and tracking to gather kinematical information about the individual electrons. SuperNEMO will improve on NEMO-3 by using a larger mass of isotope, lowering backgrounds, and improving the energy resolution. The present design is for 100 kg of \nuc{82}{Se}, but other isotopes are being considered. It will have a modular design of 20 thin-source planes of 40 mg/cm$^2$ thickness. Each source will be contained within a geiger-mode drift chamber enclosed by scintillator and phototubes. Timing measurements from digitization of the scintillator and drift chamber signals will provide topological information such as the event vertex and particle directionality. The modules will be surrounded by water and passive shielding. A one-module demonstrator with 7 kg of \nuc{82}{Se} is planned to be in operation by 2014. This Demonstrator will have only passive shielding. The complete experiment will be ready by the end of the decade in an extension of the LSM Modane in the Frejus Tunnel in France.

\begin{table}[h]
\tbl{A summary list of the \BBz\ proposals and experiments.}
{\begin{tabular}{@{}cccccc@{}} \toprule
Experiment				&   Isotope		& Mass		&  Technique										& Present Status			& Location  \\
\colrule
AMoRE\cite{Kim11,Lee11}		&\nuc{100}{Mo}	& 50 kg		&CaMoO$_4$ scint. bolometer crystals			& Development			& Yangyang \\
CANDLES\cite{Kis09}		&  \nuc{48}{Ca}	& 0.35 kg	& CaF$_2$ scint. crystals						& Prototype 				&  Kamioka \\
CARVEL\cite{zde05}		&\nuc{48}{Ca}	& 1 ton		& CaF$_2$ scint. crystals						& Development			& Solotvina        \\
COBRA\cite{zub01}		&  \nuc{116}{Cd}	& 183 kg		& \nuc{enr}{Cd} CZT semicond. det.				& Prototype             &  Gran Sasso   \\
CUORE-0\cite{Aless11}		&  \nuc{130}{Te}	& 11 kg		& TeO$_2$ bolometers							& Construction - 2012	&  Gran Sasso            \\
CUORE\cite{Aless11}		&  \nuc{130}{Te}	& 203 kg		& TeO$_2$ bolometers								& Construction - 2013	&  Gran Sasso            \\
DCBA\cite{ish00}			& \nuc{150}{Ne}	& 20 kg		&\nuc{enr}{Nd} foils and tracking				& Development			& Kamioka  \\
EXO-200\cite{Ack11}		&  \nuc{136}{Xe}	& 160 kg		&Liq. \nuc{enr}{Xe} TPC/scint.					& Operating - 2011		& WIPP             \\
EXO\cite{dan00a}		&  \nuc{136}{Xe}	& 1-10 t		&Liq.  \nuc{enr}{Xe} TPC/scint.						& Proposal              &  SURF             \\
GERDA\cite{sch05}		&  \nuc{76}{Ge}	& $\approx$35 kg	&\nuc{enr}{Ge} semicond. det.  				& Operating - 2011		&  Gran Sasso             \\
GSO\cite{dane01}			& \nuc{160}{Gd}	& 2 ton		&Gd$_2$SiO$_5$:Ce crys. scint. in liq. scint.					& Development 			&  \\
KamLAND-Zen\cite{Efr11}	& \nuc{136}{Xe}	& 400 kg	& \nuc{enr}{Xe} disolved in liq. scint.          		& Operating - 2011    & Kamioka \\
LUCIFER\cite{Giu10,Arnaboldi11}		& \nuc{82}{Se} & 18 kg		& ZnSe scint. bolometer crystals		& Development			& Gran Sasso		\\
\MJ\cite{Aguayo2011,Phillips2011,Schubert2011}			&  \nuc{76}{Ge}	& 26 kg		&\nuc{enr}{Ge} semicond. det. & Construction - 2013	&  SURF             \\
MOON	\cite{eji07}		&  \nuc{100}{Mo}	& 1 t		&\nuc{enr}{Mo}foils/scint.							& Development			&               \\
SuperNEMO-Dem\cite{Arn10}	&  \nuc{82}{Se}	& 7 kg		& \nuc{enr}{Se} foils/tracking				& Construction - 2014       &  Fr\'{e}jus             \\
SuperNEMO\cite{Arn10}	&  \nuc{82}{Se}	& 100 kg		& \nuc{enr}{Se} foils/tracking					& Proposal - 2019       &  Fr\'{e}jus             \\
NEXT	\cite{Gom11,Yah10}		&  \nuc{136}{Xe} & 100 kg	& gas TPC                                    	& Development - 2014     &   Canfranc            \\
SNO+\cite{che05,Chen2008}			&  \nuc{150}{Nd} & 55 kg		& Nd loaded liq. scint.                       	& Construction - 2013    &  SNOLab             \\
\hline
\end{tabular}\label{tab:FutureExperiments} }
\end{table}%

\section{The Number of Required Experimental Results}
\label{sec:NumberExpts}
Although the existence of \BBz\ would prove that neutrinos are massive Majorana particles\cite{sch82}, it is not clear which lepton violating process might actually mediate the decay. Since we now know that light massive neutrinos do exist and would mediate \BBz\ if they are Majorana particles, that is the simplest model that could incorporate \BBz. It is this hypothesis that is expressed in Eqn.~\ref{eq:rate}. Nevertheless, other physics might be present including: heavy Majorana neutrino exchange, right-handed currents (RHC), and exchange mechanisms that arise from R-Parity violating supersymmetry (RPV SUSY) models.

In contrast to Eqn.~\ref{eq:rate}, the \BBz\ rate can be written more generally: 

\begin{eqnarray}\label{eq:BBRate}
[T^{0\nu}_{1/2}]^{-1} =G^{0\nu} |M_{0\nu} \eta|^{2}
\end{eqnarray}

\noindent where $\eta$ is a general lepton number violating parameter (LNVP) that was previously given by \mee.   The LNVP $\eta$ contains all of the information about lepton number violation.

The LNVP takes on different forms for different \BBz\ mechanisms. In addition, \Mz\ may also depend on the mechanism. The heavy-particle models represent a large class of theories that involve the exchange of high-mass ($>$1 TeV) particles.  For example, leptoquarks\cite{Hir96b} have very similar \Mz\ to RPV SUSY\cite{Hir96}. Left-right symmetric models can lead to right-handed current models\cite{Doi85} or heavy neutrino exchange models\cite{Hir96b}. Scalar bilinears\cite{kla03c} might also mediate the decay but explicit matrix elements have not been calculated yet. For SUSY and left-right symmetric models, effective field theory\cite{Pre03} has been used to determine the form of the effective hadronic operators from the symmetries of the \BBz-decay operators in a given theory. In all of these cases the estimates of \Mz\ are not as advanced as that for light neutrino exchange and more work is needed.

A number of authors\cite{Geh07,Dep07,fog09,sim10,fae11,Ali2010} have tried to estimate the number of measurements required to discern the underlying physics. The assumption that is critical to these arguments is that the spread in \Mz\ due to different models reflects the true variation, which is clearly speculative given the uncertainties still remaining in those calculations. With this assumption in mind, the conclusion is that 3 or more experiments with a precision of about 20\% for both experiment and theory are needed to have any hope of discerning the underlying physics. There are other motivations that require multiple \BBz\ experimental results\cite{Ell06}. The need to prove that the observation is indeed \BBz\ and not an unidentified background is not the least among them. Therefore, a general conclusion from these papers, and those similar to them, indicate that at least 3 (and very likely 4) \BBz\ experiments along with significant theoretical effort are warranted. The need for and utility of several precision experimental results is the critical conclusion.

\section{Toward the Normal Hierarchy Neutrino-Mass Scale}
Atmospheric neutrino oscillation results indicate that at least one neutrino mass eigenstate is about 50 meV or more. In the inverted hierarchy with a small value for the lightest $m_i$ (\ml), \mee\ would be about 20-50 meV. At this value for \mee\ the \BBz\ half life would be near $10^{27}$ y. It is this scenario that motivates the size of the tonne-scale experiments that will evolve from the projects listed in Section~\ref{sec:expt}.
In the normal hierarchy, when \ml\ is near 0, \mee\ is 5-10 meV. To reach such sensitivity, \BBz\ experiments must be capable of observing a half-life near $10^{29}$ y. This will require 100 tonnes of isotope. The technology for enriching isotopes at this quantity is not yet cost effective and research will be required to develop it. Furthermore, energy resolution will be critical to make sure that background from the tail of the \BBt\ spectrum doesn't mask the \BBz\ signal due to its much faster decay rate. Techniques to fabricate a very large number of high-resolution crystal detectors in a cost effective way must be developed or, alternatively, improving the resolution of large scintillator detectors is required. At any mass scale, once \BBz\ is observed, technologies that can measure the full kinematic information of the process will be a critical component of the \BB\ program.

The various \BB\ experiments prior to 2010 used up to $\sim$10 kg of isotope and were successful in measuring \BBt\ in $\sim$10 isotopes and demonstrated sensitivity to a half-life for \BBz\ near $10^{25}$ y. In the upcoming years until 2015, experiments using up to a few hundred kg will built and operated as discussed in Section~\ref{sec:expt}. From these experiments we anticipate a robust test of the recent claim and either limits or measurements near 100 meV or better. At that point in time, there will be a decision point depending on whether \BBz\ is convincingly detected or not. If it is observed, the program will proceed toward a large number of experiments sensitive to the appropriate \mee\ scale in an effort to elucidate the underlying physics as discussed in Section~\ref{sec:NumberExpts}. If not, then the program will push on to the tonne-scale and atmospheric mass-scale sensitivity below 50 meV followed by another decision point. Depending on whether \BBz\ is observed or not, a collection of experiments will be developed the better to study the decay or the program will aspire to even greater sensitivity - reaching toward the normal hierarchy mass scale. 

Reference \refcite{Tretyak2011} relates a cautionary tale about double beta decay experiments. It describes a number of past \BB\ efforts that produced, in some cases, incorrect conclusions, and in others, enticing indications of new physics that took time to understand as background or instrumental effects. Although the article paints some past experimental efforts with a overly negative broad brush, it does make clear that as we reach for ever greater sensitivity, we are likely to encounter new background effects with which we are presently unfamiliar. Any claim for \BBz\ will take time to confirm and fully understand.

\section{Conclusions}
The worldwide research program in \BB\ is making fast progress due to the great interest in the subject. This interest arises because we now know that neutrinos have mass and because \BBz\ is the only practical way to discover if the neutrino is its own anti-particle. In fact, \BBz\ can only exist if neutrinos are massive Majorana particles. Experimental technologies have now advanced to the point that sensitivity to the inverted hierarchy mass scale will soon be achieved. It is also now possible to imagine reaching the normal hierarchy scale even if the concepts aren't yet feasible. However, to interpret a measured decay rate (or limit) in terms of constraints on neutrino parameters requires an understanding of the matrix elements. \BB\ is a second order weak process that proceeds through many possible intermediate states and therefore the matrix element calculation is a difficult theoretical challenge. Not only has the theory made great recent strides in developing techniques to calculate \Mz, a number of experimental efforts are investigating the nuclear physics involved to help better understand these calculations. All told, this is a very exciting time for the study of \BB.

\section*{Acknowledgments}

 I gratefully acknowledge the support of the U.S. Department of Energy through
the LANL/LDRD Program for this work.  I also gratefully acknowledge the support of the U.S. Department of Energy, Office of Nuclear Physics under
Contract No. 2011LANLE9BW. I wish to thank Hamish Robertson and Frank Avignone for a careful reading of this manuscript and many useful suggestions. I thank many of my experimental colleagues who read the sections on their respective experiments and improved the descriptions. These include F. T. Avignone III, M. Chen, Yu. Efremenko, G. Gratta, K. Lang, D. Nygren and S. Schonert.

\bibliographystyle{iopart-num}
\bibliography{DoubleBetaDecay.bbl}
\end{document}